\documentclass[angeo]{copernicus}
\usepackage{graphicx}
\usepackage{amsmath}
\usepackage{epstopdf}

\begin{document}

\title{Non-adiabatic electron behaviour due to short-scale electric field structures at collisionless shock waves}

\author{V. See}
\author{R. F. Cameron}
\author{S. J. Schwartz}
\affil{Blackett Laboratory, Imperial College London, London, SW7 2AZ, UK}

\runningtitle{Electron behaviour due to short-scale structures at shock waves}

\runningauthor{V. See, R. F. Cameron \& S. J. Schwartz}

\correspondence{V. See (victorwcsee@gmail.com)}

\received{}
\revised{}
\accepted{}
\published{}

\firstpage{1}

\maketitle

\begin{abstract}
Under sufficiently high electric field gradients, electron behaviour within exactly perpendicular shocks is unstable to the so-called trajectory instability. We extend previous work paying special attention to shortiscale, high amplitude structures as observed within the electric field profile. Via test particle simulations, we show that such structures can cause the electron distribution to heat in a manner that violates conservation of the first adiabatic invariant. This is the case even if the overall shock width is larger than the upstream electron gyroradius. The spatial distance over which these structures occur therefore constitutes a new scale length relevant to the shock heating problem. Furthermore, we find that the spatial location of the short-scale structure is important in determining the total effect of non-adiabatic behaviour - a result that has not been previously noted.

\end{abstract}

\introduction 
Collisionless shockwaves occur throughout the universe.  While often cited as the production source for high energy cosmic rays, the heating mechanisms that act on the different sub-populations of particles are still not entirely understood. Numerous studies have been conducted into the electron heating problem, with the characteristic scale length of the shock emerging as an important parameter governing the evolution of the electron distribution \citep{RefWorks:8, RefWorks:39, RefWorks:43}. Additionally, despite the amount of work conducted on shock scale lengths, there is still a lack of consensus regarding the relative scales over which the magnetic and electric fields in shocks actually vary.

Electrons are expected to behave adiabatically, conserving their magnetic moments $\mu_{m} \equiv W_{\perp}$/$B$, as long as the shock width is larger than the upstream electron gyroradius. This behaviour allows an electron to change the kinetic energy associated with its gyrovelocity perpendicular to the magnetic field smoothly as it crosses the shock. However,  \citet{RefWorks:6} showed that in the presence of an electric field with constant gradient,
\begin{equation}
	\mathbf{E}=\left(E_0+\frac{\partial E_x}{\partial x} x\right)\mathbf{\hat{x}},
\end{equation}
particles will gyrate at an effective frequency, given by

\begin{equation}
	\Omega^{2}_{eff}=\Omega^2-\frac{q}{m}\frac{\partial E_x}{\partial x},
	\label{egEffGyR}
\end{equation}
where $\Omega_{eff}$ and $\Omega$ are the effective and normal gyrofrequencies, $q$ is the charge on the particle and $m$ is the particle mass. The effective gyrofrequency must then be used in calculating the gyroradii, i.e.

\begin{equation}
	r^{eff}_g=\frac{v}{\Omega_{eff}},
\end{equation}
where $r^{eff}_g$ is a new effective gyroradius and $v$ is the gyrovelocity of the particle. The condition for adiabatic behaviour must be revised such that the shock width is much bigger than the effective gyroradius. Equation (\ref{egEffGyR}) shows that, for certain values of $\partial E_x$/$\partial x$, the effective gyrofrequency can approach zero corresponding to an extremely large effective gyroradii. Non-adiabatic electron behaviour is therefore possible, even at shocks with scale lengths much larger than an upstream gyroradius.

The link between scale lengths and non-adiabatic heating was explored by \citet{RefWorks:7}. The authors conducted a theoretical analysis of electron trajectories at exactly perpendicular shocks and identified the so-called trajectory instability. This instability causes two neighbouring electron trajectories to diverge exponentially from each other in phase space, causing a breaking of magnetic moment conservation, wherever $\Omega^2_{eff} < 0$, i.e. as long as the following instability criterion is obeyed:

\begin{equation}
	-\frac{e}{m}\frac{\partial E_{x}}{\partial x}-\Omega^{2}>0,
	\label{eqInCrit}
\end{equation}
where $e$ is magnitude of the electronic charge, $m$ is the electron mass, $\Omega$ is the electron gyrofrequency and $\partial E_{x}$/$\partial x$ is the electric field gradient along the shock normal. The criterion requires that the electric field gradient be above some critical value or, equivalently for a given cross-shock potential, that the scale length of the electric field be below some critical value. This can be fulfilled even if the upstream gyroradius is smaller than the shock scale. The authors then showed via a series of test-particle simulations that the onset of the trajectory instability coincides with the onset of non-adiabatic heating. While \citet{RefWorks:7} draw a strong connection between the scale length of the shock and subsequent heating, they do not alter the scales of the electric and magnetic fields independently of each other, nor do they study the effect of displacing one with respect to the other.

This work was subsequently extended into the oblique regime by \citet{RefWorks:8}. In this paper the authors also included terms that account for the changing magnetic field, which were previously neglected, and found that the divergence in phase space always occurs and that the rate of divergence is dependent on the gradients of both the magnetic and electric fields. 

Further relevant work is done by \citet{RefWorks:26}. Two approaches were used to analyse the demagnetisation of the electrons at the shock front. In the first instance, nonstationary and nonuniformity effects were included in the form of a full-particle self-consistent simulation whilst in the second instance these effects have been removed. The authors found that the fraction of electrons which become demagnetised depends on the nonstationary behaviour found at shocks. However, it is difficult to attribute this result to any particular process or feature of the shock since it is impossible to systematically vary particular variables of interest in a full particle code.

It is clear that the relative scales over which the magnetic and electric fields vary have a large impact on the type of electron heating that occurs. Indeed, the relative field scales of shocks is a topic which we study within this paper. In their paper, \citet{RefWorks:8} outlined possible relationships between the fields, though it is a matter of contention which of them occurs in reality, since various observations and simulations support differing views. 

It is common that both scales have the same order of magnitude in simulations and observations \citep{RefWorks:7, RefWorks:12, RefWorks:13, RefWorks:15, RefWorks:1, RefWorks:17, RefWorks:18, RefWorks:19, RefWorks:20}. Balikhin and Gedalin (1994) suggest that the variation of electron heating with upstream electron thermal Mach number $v_{flow}$/$v_{thermal-e}$, reported by \citet{RefWorks:5}, can be recovered in this simple configuration. On the other hand, \citet{RefWorks:22} analysed a shock where the electrostatic potential varied over a scale larger than the magnetic field ramp. 

However, \citet{RefWorks:9} reported on so-called iso-magnetic jumps which were observed in laboratory plasma experiments whilst \citet{RefWorks:10} reported the observations from ISEE-1 of large changes in the electric field over scales much shorter than the magnetic field ramp. More recently, \citet{RefWorks:24} and \citet{RefWorks:25} have shown the existence of short-scale, high-amplitude electric field structures or 'spikes' within the overall electric field profile with \citet{RefWorks:25} speculating that the spikes in the electric field profile may lead to incoherent heating of the electrons. 

In this paper, we will show for the first time that this is indeed possible. Using test-particle simulations, we will find the effect of varying the electric field scale length independently of the magnetic scale length; which has not been done before. Additionally, we will vary the location of the electric field within the shock. We also investigate the consequences of an electric field spike within the shock. In doing so, we will demonstrate that these electric field spikes constitute a new scale length which is important to the shock heating problem, and that its location within the shock layer can dramatically change the amount of heating observed.

The rest of this paper will be structured as follows. Section 2 will cover the details behind the simulation, with the results and analysis following in section 3. Conclusions follow in section 4.

\section{The Simulation}
\subsection{Field Profiles}
\begin{figure}[h]
	\begin{center}
	\includegraphics[width=\columnwidth]{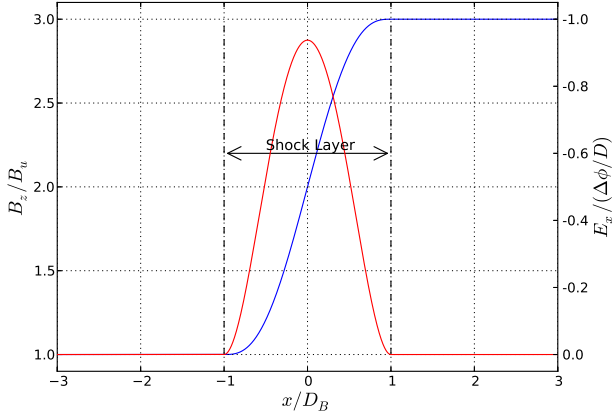}
	\end{center}
	\caption{Profiles for the electric, $E_{x}$ (red curve), and magnetic, $B_{z}$ (blue curve), fields in dimensionless units (see Eq.~(\ref{eqExField}) - (\ref{eqBzField})). The shock width is set by the parameter $D_B$, with the $D_B=D_E=1$ case illustrated in the figure. Note that the $E_x$ scale is negative.}
	\label{figFieldProfile}	
\end{figure}

A test-particle approach, where static electromagnetic fields are prescribed, is chosen for this investigation. The normalisation details can be found in \citet{RefWorks:7} and are briefly reproduced here. Time is normalised to the inverse gyrofrequency, $\Omega^{-1}$; coordinates are normalised to the electron inertial length, $c\omega_{pe}^{-1}$; velocity is normalised to the upstream Alfv\'{e}n speed, $v_A$; and magnetic fields are normalised in terms of the upstream magnetic field strength, $B_u$. The field profiles used are based upon the profiles described by \citet{RefWorks:7}. They are idealised versions of exactly perpendicular collisionless shocks. The field profiles are shown in Fig.~\ref{figFieldProfile} and are given by Eq.~(\ref{eqExField}), (\ref{eqEyField}) and (\ref{eqBzField}). The shock is at rest in the simulation frame, with the upstream-pointing normal in the $-\hat{x}$ direction.

\begin{equation}
	E_{x}=-\frac{15\Delta\phi_{0}}{16D_E}\left(\left(\frac{x}{D_E}\right)^{2}-1\right)^{2} 
	\label{eqExField}
\end{equation}
\begin{equation}
	E_{y}=M_A
	\label{eqEyField}
\end{equation}
\begin{equation}
	B_{z}=2+\frac{1}{8}\left(3\left(\frac{x}{D_B}\right)^{5}-10\left(\frac{x}{D_B}\right)^{3}+15\left(\frac{x}{D_B}\right)\right)
	\label{eqBzField}
\end{equation}
\begin{figure*}[t]
	\begin{center}
	\includegraphics[width=15cm]{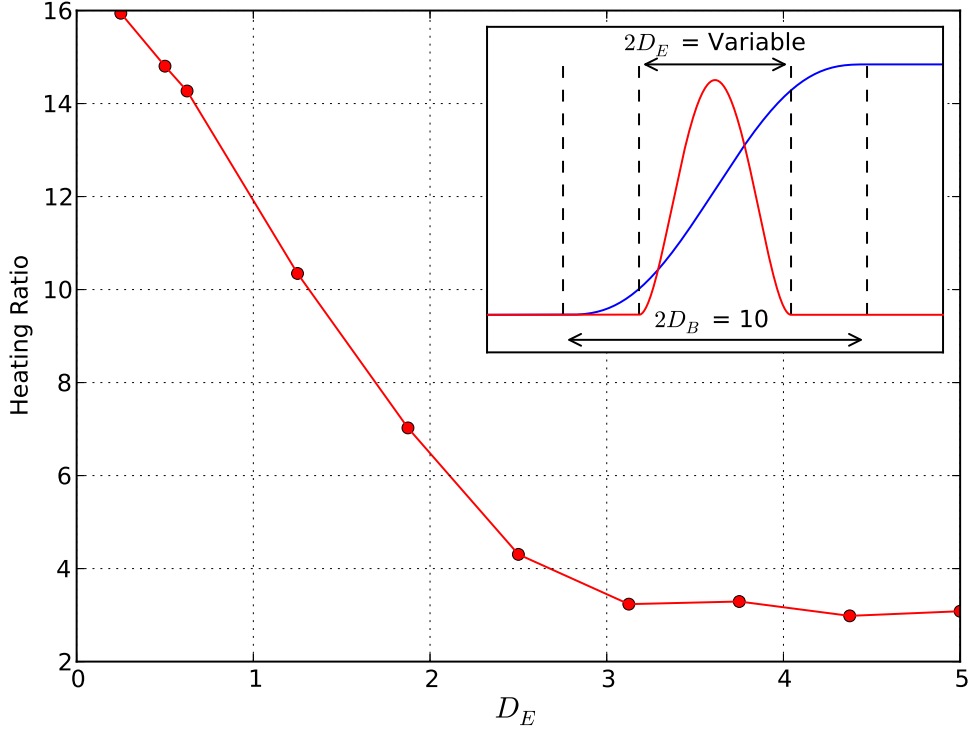}
	\end{center}
	\caption{Ratio of downstream to upstream electron temperature as a function of electric field scale length, $D_E$. The magnetic field scale length is kept fixed at $D_B=5$. A sketch of the field profiles and their relative scale lengths is shown in the inset. For large $D_E$, the heating stays adiabatic as the electric field scale decreases. Once the adiabaticity is broken at scale lengths shorter than roughly $D_E=3$, however, there is a negative correlation between the heating ratio and the electric scale length.}
	\label{figVaryEWidth}
\end{figure*}

\noindent Here, $\Delta\phi_{0}$ is the cross-shock potential and chosen to be 300\unit{eV} unless stated otherwise. $E_y$ is constant everywhere and calculated from the upstream bulk electron velocity and magnetic field strength, $E_y=V_uB_u$. We use values of $V_u=400$\unit{kms^{-1}} and $B_u=5$\unit{nT} which are typical for earth's bow shock. When normalised, $E_y$ is equal to the Alfv\'{e}nic Mach number, $M_A$, which we choose to be Mach 8.  $D_E$ and $D_B$ are the half-electric and half-magnetic field widths normalised to the electron inertial length. Equation (\ref{eqExField}) only applies within the region of space $-D_E>x>D_E$. Everywhere outside this region, $E_x=0$. Similarly, Eq.~(\ref{eqBzField}) only applies within $-D_B>x>D_B$, taking the values $B_{z}=1$ for $x<-D_B$ and $B_{z}=3$ for $x>D_B$. Adiabatic electron behaviour, conserving magnetic moment, would therefore correspond to a three-fold increase in the temperature of the electron distribution based on the jump in the magnetic field. We have chosen to use two scale lengths, $D_E$ and $D_B$, rather than the single parameter, $D=D_E=D_B$, that \citet{RefWorks:7} use because it is important for this study that we are able to vary the two scale lengths independently. These particular forms were chosen by \citet{RefWorks:7} because they are smooth and well behaved at the shock edges and throughout the shock layer.

\subsection{Electron Distribution}
For each simulation run, a Maxwellian distribution at a temperature of 10\unit{eV} consisting of 600 electrons is initialised far upstream from the shock. Since the shock is exactly perpendicular, the electrons only require two degrees of freedom in velocity space allowing us to set $v_z=0$.

For the purposes of this investigation, the temperature corresponding to the two perpendicular ($x$, $y$) degrees of freedom will be defined as follows:

\begin{equation}
	T=\frac{m}{2k_B}\left\langle \left(\mathbf{v-\left\langle \mathbf{v}\right\rangle }\right)^{2}\right\rangle,
\end{equation}
i.e. the temperature is proportional to the variance of the velocity vectors of all the electrons in the distribution. In practice, the parameter that will be of interest is the heating ratio, $R_{H}$; that is the ratio of the far downstream electron distribution temperature to the far upstream temperature. 

\begin{figure*}[t]
	\begin{center}
	\includegraphics[width=15cm]{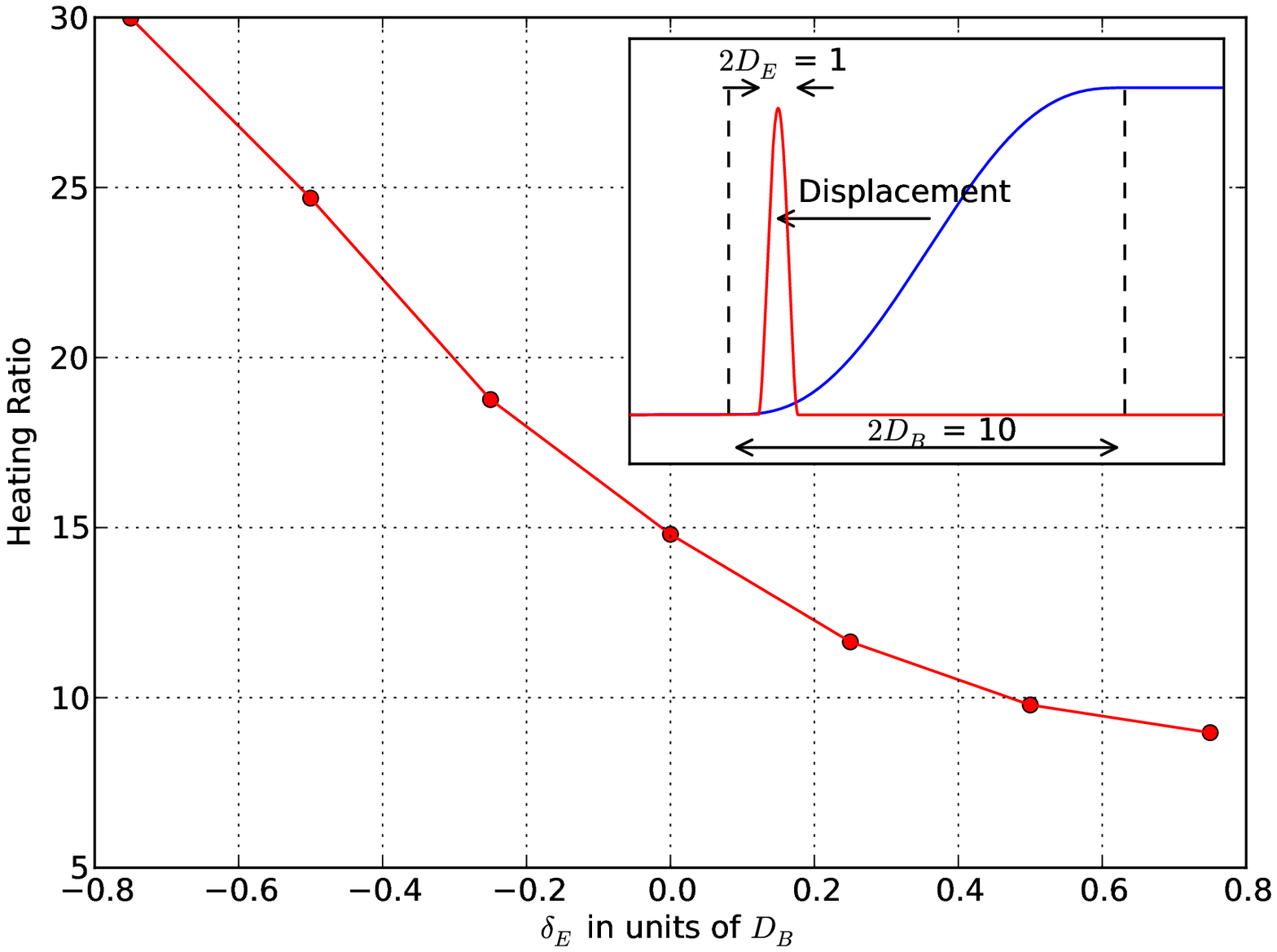}
	\end{center}
	\caption{Ratio of downstream to upstream electron temperature as a function of electric field displacement, $\delta_E$. The electric and magnetic field scale lengths are kept at $D_E=0.5$ and $D_B=5$. A sketch of the field profiles and their relative scale lengths is shown in the inset. The displacement of the electric field spike given in terms of $D_B$, i.e. $\delta_E=-1$ corresponds to the center of variations in the electric field coinciding with the upstream edge of the shock layer. The heating ratio is greater for displacements towards the upstream edge of the shock. Conversely, when the electric field is displaced towards the downstream end, the heating ratio is lower.}	
	\label{figVaryEDisp}
\end{figure*}

\section{Results and Analysis}
To investigate short-scale electric field structures, it will be instructive to investigate, separately, the scale and location of the cross-shock electric field, $E_x$. We will then move onto a final set of simulations in which the cross-shock electric field will vary over the same scale as the magnetic field with a spike embedded within it to better represent a real shock.

\subsection{Electric Field Scale Length}
For this experiment, we will vary $D_{E}$ whilst holding $D_{B}$ and the total cross-shock potential, $e\Delta\phi_0$, fixed. The starting shock parameters that will be considered are $D_E=D_B=5$ and $e\Delta\phi_0=300$\unit{eV}. This scale length corresponds to a shock width, 2$D_B$, of 11.2 upstream gyroradii for a 10\unit{eV} electron. These conditions are adiabatic as shown in Fig.~4 of \citet{RefWorks:7} and will be the control case against which other simulations are compared. 

Figure \ref{figVaryEWidth} shows that as $D_{E}$ is decreased, the heating remains roughly adiabatic for larger $D_E$ before increasing rapidly for scale lengths below $D_{E}\thicksim3$. At these smaller electric scale lengths, the heating is significantly non-adiabatic. By holding the cross-shock potential constant and decreasing $D_{E}$, the electric field gradient becomes larger. This result should therefore not present much surprise since it is already known that the separation of the adiabatic and non-adiabatic regimes in perpendicular shocks depends on the electric field gradient as given by Eq.~(\ref{eqInCrit}). According to this criterion, the threshold of the trajectory instability occurs at $D_E\sim3.1$ for the parameters of our simulation.

\begin{figure*}[t]
	\begin{center}
	\includegraphics[width=15cm]{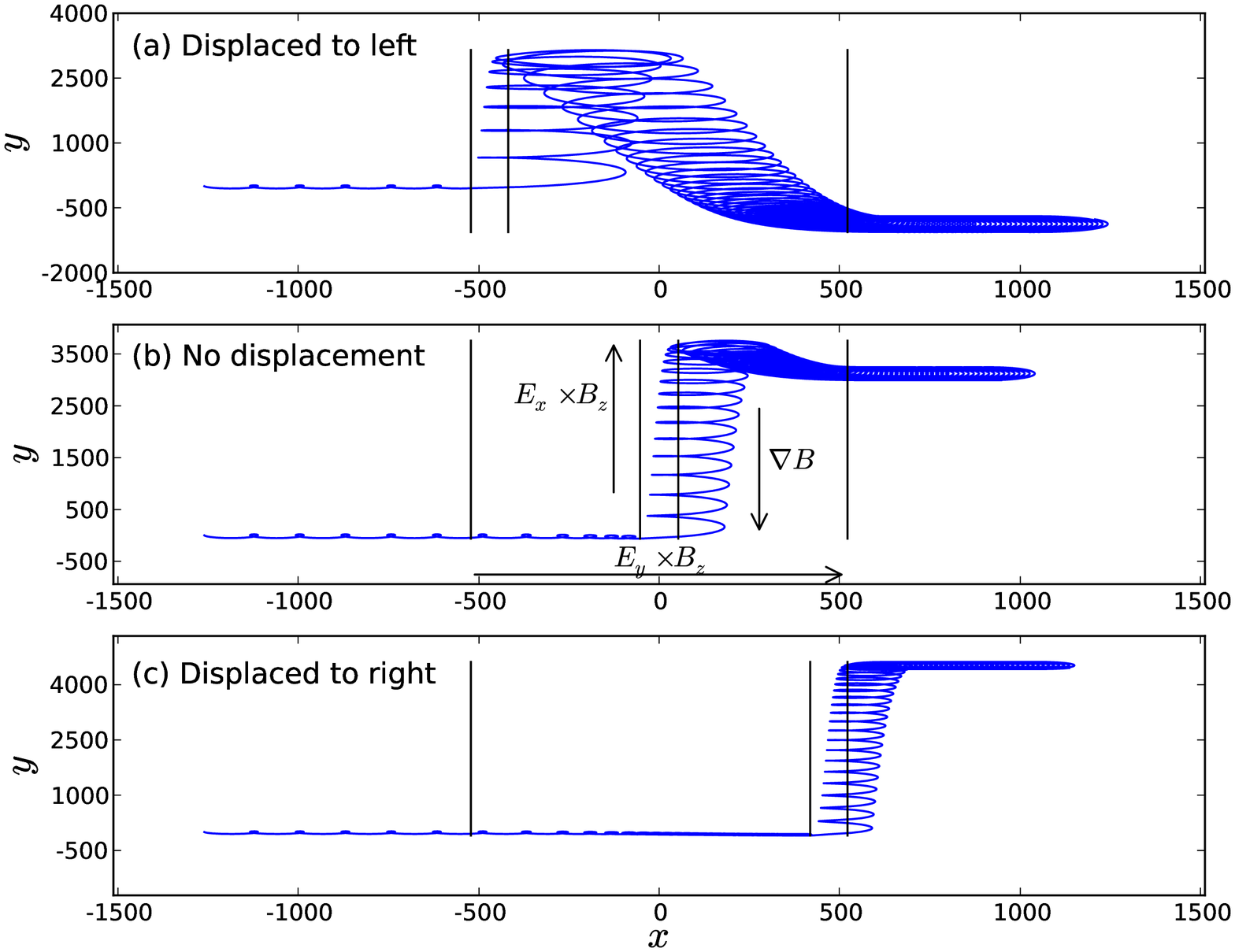}
	\end{center}
	\caption{Three electron trajectories in the $xy$ plane for different displacements of the electric field profile. In all three cases the magnetic field variations occur between the two outer vertical lines. The electric field variations are bound by the two left-most lines in panel (a), the centre two lines in panel (b), and the two right-most lines in panel (c). All other parameters are fixed. The drift directions are shown in panel (b). The panels show that when the electric field is displaced upstream, i.e.\ panel (a), the electron will drift in the negative $\hat{\textbf{y}}$ direction a lot more compared to when the displacement is downstream, i.e. panel (c)}	
	\label{figElecTraj}
\end{figure*}

\subsection{Displacement of Electric Field}
Having varied the width of the electric field profile, its position relative to the rest of the shock can be altered since $D_E$ is smaller than $D_B$. The inset of Fig.~\ref{figVaryEDisp} shows the displacement of the electric field with respect to the magnetic field such that their centres of variation no longer coincide. For this set of simulations, we fix $D_{E}=0.5$. As before, $D_{B}=5$ and $e\Delta\phi_0=300$\unit{eV}. Figure \ref{figVaryEDisp} shows a clear trend of higher (lower) heating for displacements towards the upstream (downstream) side of the shock.

To understand why displacing the electric field would change the amount of heating, despite maintaining a constant electric field gradient, it is necessary to look at the drifts in the system. For the field geometries used, the electrons experience an $E_{y}\hat{\textbf{y}}\times B_{z}\hat{\textbf{z}}$ drift in the $\hat{\textbf{x}}$ direction, together with an $E_{x}\hat{\textbf{x}}\times B_{z}\hat{\textbf{z}}$ drift and a $\nabla|\textbf{B}|$ drift which are in the $+\hat{\textbf{y}}$ and $-\hat{\textbf{y}}$ directions, respectively. The $E_{y}\hat{\textbf{y}}\times B_{z}\hat{\textbf{z}}$ drift causes the electrons to drift through the shock and gain all the potential energy associated with the $E_{x}$ field, i.e. the cross-shock potential. This is fixed by the $\Delta\phi_0$ parameter. The remaining two drifts cause the electrons to travel along the shock in opposite directions. The $\nabla|\textbf{B}|$ drift is directed such that the electrons gain kinetic energy from the motional electric field, $E_y$. Conversely, the $E_{x}\hat{\textbf{x}}\times B_{z}\hat{\textbf{z}}$ drift is directed such that the electrons lose kinetic energy to this field. The latter two drifts, in addition to the fixed cross-shock potential, therefore determine the net kinetic energy gain of the electrons as they drift through the shock \citep{RefWorks:34}.

\begin{figure*}[t]
	\begin{center}
	\includegraphics[width=15cm]{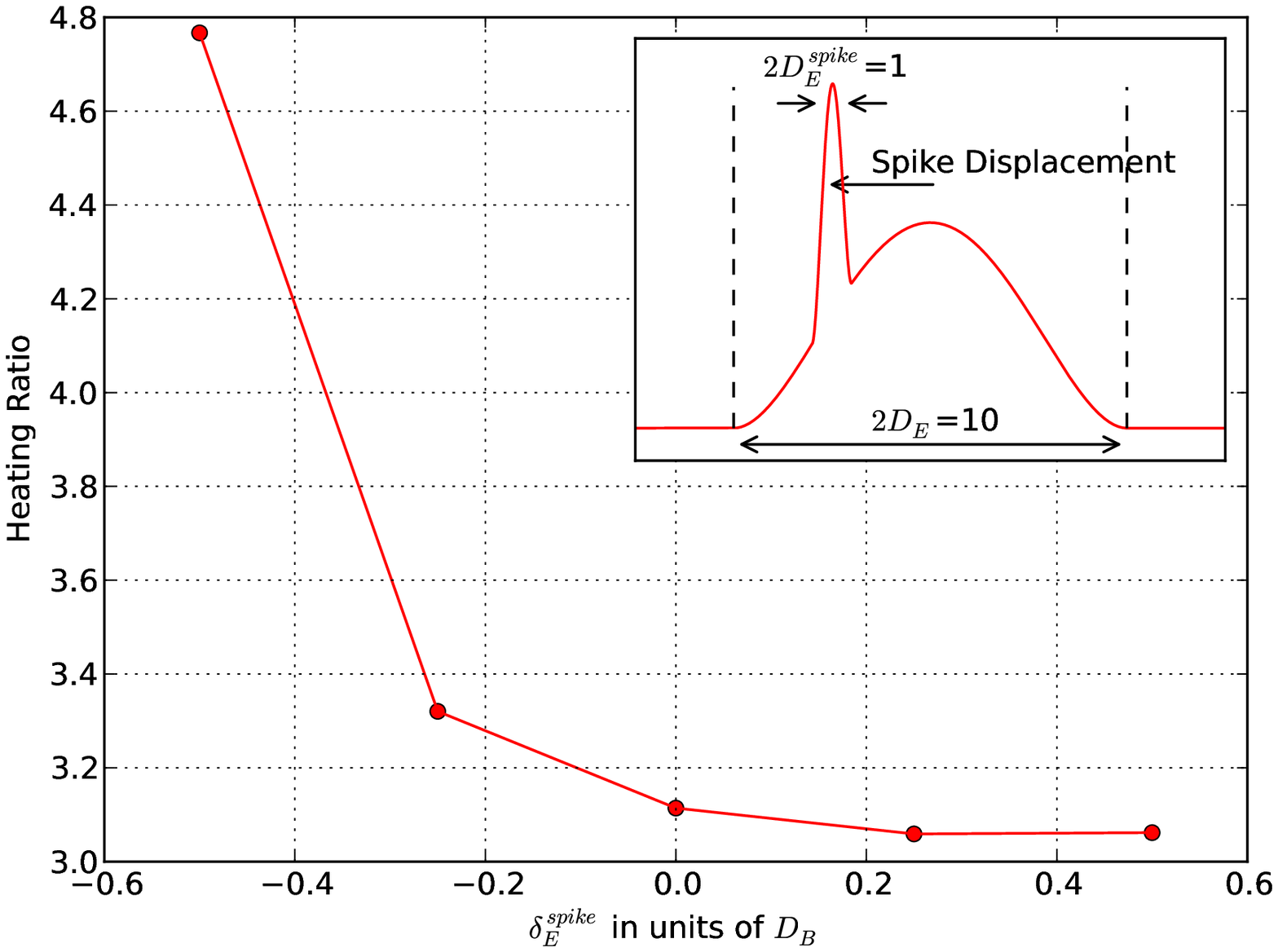}
	\end{center}
	\caption{Ratio of downstream to upstream electron temperature as a function of electric field spike displacement, $\delta^{spike}_E$. The electric and magnetic field scale lengths are kept fixed at $D_E=D_B=5$ and $D_{E}^{spike}=0.5$. A sketch of the electric field profile is shown in the inset. The magnetic field has been omitted for clarity. The displacement of the electric field spike given in terms of $D_B$ i.e. a displacement of -1 would mean that the variations in the electric field are centered exactly at the upstream edge of the shock. The heating ratio is greater for displacement towards the upstream edge of the shock. Conversely, when the electric field spike is displaced towards the downstream end, the heating ratio is lower. For all displacements, the heating is non-adiabatic.}
	\label{figSpikes}
\end{figure*}

It will be useful to compare two limiting cases in our explanation. The electrons will drift through most of the shock before encountering the electric field when it is displaced downstream. However, when the field is displaced upstream, the electrons will encounter it immediately and gain the entire cross-shock potential straight away. Since the $\nabla|\textbf{B}|$ drift speed is proportional to the kinetic energy of the electron, the magnitude of the $\nabla|\textbf{B}|$ drift will be larger in the second case as it has gained the energy from crossing the $E_{x}$ field earlier. Figure \ref{figElecTraj} shows the trajectories of three electrons which demonstrate this effect. The outer vertical lines represent the outer edges of the shock i.e. $x=\pm D_{B}$ with the inner vertical lines representing the edges of the displaced electric field, i.e. $x=\delta_{E}\pm D_{E}$ where $\delta_{E}$ is the displacement of the electric field. All parameters are kept the same with the exception of the displacement of the electric field. The electron in panel (a) immediately picks up the cross-shock potential energy, $e\Delta\phi_0$. Initially the $E_{x}\hat{\textbf{x}}\times B_{z}\hat{\textbf{z}}$ drift dominates, resulting in the loss of some of this energy. The $\nabla|\textbf{B}|$ drift then operates in the $E_x=0$ region where, due to the enhanced perpendicular velocity, a large drift velocity results in a net $-\hat{\textbf{y}}$ drift. This corresponds to a large non-adiabatic energy increase. Panel (b) is similar but the $\nabla|\textbf{B}|$ drift is less effective since the electron spends less time in the post-$E_x$ region, allowing less time for the $\nabla|\textbf{B}|$ drift to act. In panel (c), there is no space for the $\nabla|\textbf{B}|$ drift to act after the electrons have crossed the cross-shock potential. The $E_{x}\hat{\textbf{x}}\times B_{z}\hat{\textbf{z}}$ drift reduces the energy gained from $e\Delta\phi_0$ the most compared to the other panels.

Outside the region in which $E_x$ is non-zero, the energy gains associated with the $\nabla|\textbf{B}|$ drift are roughly consistent, as expected, with adiabatic compression in the increasing magnetic field. Since this multiplies the existing particle energy, it gives the most energy to trajectories suffering early non-adiabatic processes as in panel (a).

To summarise, when the electric field is displaced downstream, the electrons drift through most of the shock adiabatically, losing energy as a result of the $E_x\hat{\textbf{x}}\times B_z\hat{\textbf{z}}$ drift, before encountering the non-adiabatic divergence in phase space as discussed by \citet{RefWorks:7}. When the electric field is displaced upstream, the electrons immediately experience the phase space divergence. As the electrons drift through the rest of the shock, they will undergo a further expansion in phase space due to the $\nabla|\textbf{B}|$ drift, the magnitude of which is larger when electric field is displaced upstream. This leads to a higher heating of the electron distribution.

We note that the trajectory instability is an essential ingredient in this non-adiabatic behaviour. While the displacement of the electric field toward the upstream enhances the instability by keeping the gyrofrquency, $\Omega$, lower in Eq. (4), experiments with different values of constant $\Omega$ (not shown) are inconclusive. Thus, we prefer to discuss the non-adiabatic behaviour in terms of the various particle drifts. Other experiments (not shown) in which a field with a larger $D_E$ is displaced remain adiabatic.

\subsection{Shock Spikes}
Whilst it has been instructive to consider these simulations, observations show structures with a scale much smaller than the total shock width embedded within a larger overall electric field profile \citep{RefWorks:24}. In one particular shock crossing, Walker et al. identified three large-amplitude, small-scale structures, the largest of which had a peak magnitude of around 45\unit{mV/m}, compared to an average motional electric field of around 14\unit{mV/m}. These were the largest amongst the field disturbances observed in the shock and occurred over the middle 50\% of the shock transition. The authors estimate that the width of these structures to be around $1-5c\omega_{pe}^{-1}$, with the magnetic field ramp occurring over a scale $\sim$10 times this. It shall be the aim of the final set of simulations to encapsulate these features, if not the actual values themselves. Most importantly, for this particular shock crossing, Walker et al. report that the structures contribute 40\% of the total cross-shock potential change.

The inset of Fig.~\ref{figSpikes} shows the field profile we used to model the electric field spikes. The electric field profile shown is constructed by adding together two profiles, both described by Eq.~(\ref{eqExField}) but with different $D_E$ values. The important features of the profile are $D_{E}=D_{B}$ and $D_{E}^{spike}\ll D_{E}$. In keeping with the ratio of the scale lengths observed by \citet{RefWorks:24}, $D_{E}^{spike}$ is one tenth of $D_{E}$. While \citet{RefWorks:24} reported that the overall electric field scale is slightly larger than that of the magnetic field ramp, the two scales have been kept equal here so as to allow for the independent investigation of the spikes alone. In any case, from the work of \citet{RefWorks:7}, we would not expect that having $D_{E}>D_{B}$ would cause the heating to be non-adiabatic, as this would make the electric field gradient smaller. For simplicity, only one electric field “spike” has been modelled. Our base case will again be the adiabatic shock where $D_{E}=D_{B}=5$ and the total cross-shock potential is 300\unit{eV}. We choose 30\unit{eV} as the cross-spike potential with the rest of the shock accounting for the remaining 270\unit{eV}. We vary the position of the electric field spike to investigate its influence on the electron behaviour. Figure \ref{figSpikes} shows that for displacements towards the downstream side of the shock, the heating is non-adiabatic, but the amount of heating above the adiabatic case is small. For upstream displacements, there is a much higher non-adiabatic component to the heating. 

The conclusions of the previous set of simulations can readily be applied here. By embedding a spike into the profile, there is now a region that satisfies the instability criterion in Eq.~(\ref{eqInCrit}) when previously there was not, thus pushing the shock into the non-adiabatic regime. The same trend is noticed, with the heating ratio having high values for displacements towards the upstream end. However, the heating ratio is much smaller in comparison to the previous simulations; this should not be surprising given that the potential drop across the spike is much smaller. Just as before, the breaking of adiabaticity occurs earlier for displacements upstream, but the phase space expansion effect due to the $\nabla|\textbf{B}|$ drift is not as pronounced since the energy gains associated with the spike are smaller. 

At $2D_E\sim1c\omega^{-1}_{pe}$, the width of our spike is at the limit of the $1-5c\omega^{-1}_{pe}$ widths reported by \citet{RefWorks:24}. We conclude that in general, the presence of short-scale enhancements to the electric field can push an otherwise adiabatic shock into the non-adiabatic regime. The width of the electric field spikes therefore constitute a new scale length that is important in the study of electron heating at collisionless shocks.

We conducted a final simulation with three spikes at displacements of -0.5$D_B$, 0.0$D_B$ and 0.5$D_B$ embedded within an underlying electric field of width $2D_E=10$. Each spike, of width $2D_E^{spike}=1$, contributed 30\unit{eV} to the cross-shock potential with the underlying profile contributing 210\unit{eV} for a total cross-shock potential of 300\unit{eV}. We find the heating ratio for this set-up to be $R_H=4.45$, which is not a surprising outcome based on our previous results. Figure \ref{figSpikes} shows that the spikes at 0.0$D_B$ and 0.5$D_B$ have a minimal effect above adiabatic electron behaviour. The non-adiabatic behaviour found here is due predominantly to the upstream-displaced spike at $-0.5D_B$. 

\conclusions
It has been the aim of this paper to look at the effect that the electric and magnetic field scales have on electron heating at collisionless shock waves with a focus on short-scale high-amplitude structures in the electric field. Our work builds on the existing work of \citet{RefWorks:7}  and is motivated by the observations of short-scale electric field structures observed by \citet{RefWorks:24} and \citet{RefWorks:25}. \citet{RefWorks:7} showed that shorter scale lengths can lead to incoherent electron heating by satisfying an instability criterion with the short-scale electric field spike observations, providing a possible means of satisfying this criterion in reality. We have shown that the presence of small-scale structures can indeed push the heating of the electron distribution from the adiabatic into the non-adiabatic regime. Specifically, the main results of this report can be summarised as follows: 

[1] Shorter-scale electric fields lead to non-adiabatic electron behaviour.

[2] The position of these electric fields has been shown, for the first time, to play an important role in determining the level of non-adiabaticity, with higher non-adiabatic behaviour observed for upstream displacements. This is due to the earlier energy gain of the electrons allowing for a large magnitude of subsequent $\nabla|\textbf{B}|$ drift. Equivalently, the magnetic moment of the electrons is increased more significantly for upstream displacements, allowing the electron to gain energy adiabatically in the subsequent magnetic field increase.

[3] This is true even when considering smaller-amplitude spikes embedded within a larger-scale electric field profile, provided that the magnitude of the electric field gradient is large enough. Such spikes have been observed \citep{RefWorks:24, RefWorks:25}.

[4] The existence, scale and location within the shock of electric field spikes are therefore important new factors to consider in the context of electron heating.

The next step would be to extend our work into the oblique regime. The new scale length associated with the electric field spikes is relevant to the discussion of shock scale lengths and heating at oblique shocks by \citet{RefWorks:8}. Other shock features of interest which could influence the electron dynamics include foot and overshoot regions, as well as the time dependence of the field profiles and higher-frequency fluctuations \citep{RefWorks:29, RefWorks:42}. Finally, it would also be interesting to study the electrons within the shock layer, where the strong trajectory instability and short scales involved might be expected to break the gyrotropy of the distributions.

\bibliographystyle{copernicus}
\bibliography{Project}{}

\end{document}